# Disentangling longitudinal and transverse modes of the $\phi$ meson through dilepton and kaon decays

In Woo Park[1,*], Hiroyuki Sako[2,†], Kazuya Aoki[3,‡], Philipp Gubler[2,§] and Su Houng Lee[1,∥]

[1]*Department of Physics and Institute of Physics and Applied Physics, Yonsei University, Seoul 03722, Korea*
[2]*Advanced Science Research Center, Japan Atomic Energy Agency, Tokai, Naka, Ibaraki 319-1195, Japan*
[3]*KEK, High Energy Accelerator Research Organization, Tsukuba, Ibaraki 305-0801, Japan*

Angular distributions of $\phi$ meson decay amplitudes of $e^+e^-$ and $K^+K^-$ channels are computed using both specific interaction Lagrangians and simple arguments relying on angular momentum conservation. Based on the obtained results, we assess methods to experimentally disentangle the longitudinal and transverse polarization modes of the $\phi$ meson and discuss advantages and disadvantages of employing either the leptonic or hadronic decay modes for this task.

## I. INTRODUCTION

It has long been recognized that longitudinal and transverse modes of vector fields can behave differently in a hot and/or dense medium (see, for instance, Ref. [1] and the references cited therein for discussions within thermal field theory). This difference is caused by the presence of the static medium reference frame, which distinguishes the vector mesons with spin pointing either in the direction orthogonal to or along its motion. Hence, while longitudinal and transverse spectral functions are degenerate for vector mesons at rest with respect to the medium, they can generally behave differently with increasing momentum. For vector mesons in nuclear matter, this has been explicitly demonstrated both in QCD sum rule [2] and effective theory approaches [3]. More recently, it was furthermore shown that the dispersion relation of the $\phi$ meson in nuclear matter can be parametrized using a momentum dependent "effective mass", which depends on the polarization direction with respect to the $\phi$ momentum [4,5]. One goal of the present work is to clarify how the two polarizations of the $\phi$ meson can be distinguished and possibly disentangled from the angular distributions of its leptonic ($e^+ + e^-$) and hadronic ($K^+ + K^-$) two-body decays. While the main focus will be the $\phi$, we note that many aspects of the formalism presented here can be applied to other vector mesons with minimal modifications.

Hadronic polarization phenomena have in recent years also attracted attention especially since the experimental observation of the $\Lambda$ baryon polarization, usually defined with respect to the direction perpendicular to the event plane, in noncentral heavy-ion collisions at STAR [6,7], which provided evidence for the presence of strong vorticity in such collisions. Similar measurements followed at the same and other facilities [8,9], were extended to other baryon species [10] and very recently to vector particles [11]. While many theoretical works are currently trying to understand the mechanisms for the various degrees of polarizations of the different channels, we here focus on the properties of the two polarization modes of vector mesons. Hence, our discussion will not depend on a specific production mechanism or whether the vector meson is polarized or not (an unpolarized vector meson would carry transverse and longitudinal components with probabilities of $\frac{2}{3}$ and $\frac{1}{3}$, respectively). We note, that the transverse and longitudinal modes to be discussed in this work are defined with respect to the momentum of the $\phi$ meson. As we will show, the angular dependence in the forward direction is dominated by the transverse component in the $e^+e^-$ channel, while for $K^+K^-$ the longitudinal mode will be dominant. Furthermore, the spectrum in the perpendicular direction is purely transverse in the $K^+K^-$ channel. All these features can be understood from simple arguments using conserved angular momentum and rotations making use of Wigner $D$ matrices as we can see in [12,13]. This suggests that it could be crucial to perform complementary measurements using both the dilepton and pseudoscalar meson decay channels, which could make it possible to disentangle the two polarization modes in a realistic measurement.

*darkzero37@naver.com
†hiroyuki.sako@j-parc.jp
‡kazuya.aoki@kek.jp
§gubler@post.j-parc.jp
∥suhoung@yonsei.ac.kr

Indeed, J-PARC E16 just started its operation in 2020 [14] to investigate in-medium spectral change of vector mesons using pA reactions. Vector mesons are reconstructed through $e^+e^-$ decay. The experiment could access polarization dependent spectral change utilizing the angular dependence. Furthermore, J-PARC E88 was proposed to measure $\phi \to K^+ + K^-$ decays with kaon identification detectors in the E16 spectrometer in pA reactions [15]. Its full coverage of the kaon decay angles is suitable for the invariant mass analysis depending on the $\phi$ polarization. Complementary studies for in-medium modification of $\phi$ mass spectra are expected with $\phi \to e^+ + e^-$ measurements at E16.

The paper is organized as follows. In Sec. II, we outline our calculation of the general angular distributions for the $e^+ + e^-$ and $K^+ + K^-$ decay channels of the $\phi$ meson. Physical interpretations and consequences of our findings are discussed in Sec. III, after which a summary and conclusions are given in Sec. IV. Detailed calculations are presented in the appendixes.

## II. $\phi$ MESON DECAY RATE

In this section, we consider the aforementioned decay channels $\phi \to K^+ + K^-$ and $\phi \to e^+ + e^-$ in detail, and demonstrate how the angular distribution of the decay products can be used to distinguish between the transverse and longitudinal polarization modes of the $\phi$ meson.

To compute the decay amplitudes, we use a phenomenological coupling model between the $\phi$ and $K$ meson pair and employ the vector meson dominance (VMD) model [16,17] for the coupling of the $\phi$ to the electron positron pair. The tree diagrams for the respective decay processes are given in Fig. 1. Figure 2 depicts the two body decay of the $\phi$ meson in its rest frame where the polar angle $\theta$ and the azimuthal angle $\varphi$ of the positively charged daughter particle are defined with respect to the direction of the $\phi$ meson momentum in the Lab frame, which is represented by a blue arrow. The $\phi$ meson rest frame can be constructed by rotating the Lab frame such that the z-axis is parallel to the $\phi$ meson momentum and boosting to the rest frame. The two-body partial decay width of the $\phi$ meson for each channel is given as

$$\Gamma = \frac{1}{8\pi} \frac{|\boldsymbol{p_1}|}{m_\phi^2} |\mathcal{M}|^2$$
$$= \begin{cases} \phi \to K^+ + K^- \; 4.249 \times (0.492) \text{ MeV}, \\ \phi \to e^+ + e^- \; 4.249 \times (2.974 \times 10^{-4}) \text{ MeV}, \end{cases} \quad (1)$$

where $|\boldsymbol{p_1}|$ is the momentum of both daughter particles in the center of mass (c.m.) frame and $\mathcal{M}$ is the invariant decay amplitude. The numbers in the brackets are the respective branching ratios given by the particle data group [18].

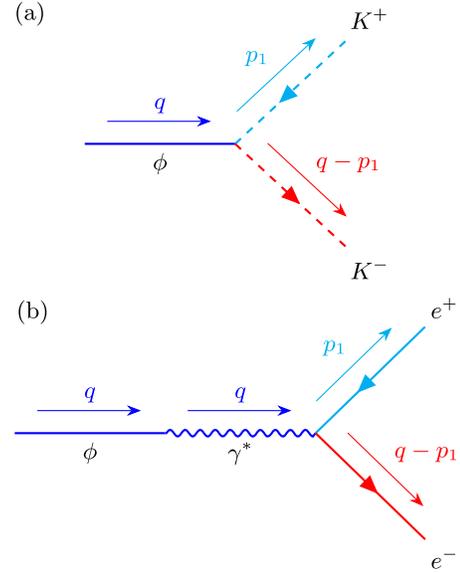

FIG. 1. Tree-level diagram of $\phi$ meson decay (a) $\phi \to K^+ + K^-$ (b) $\phi \to \gamma^* \to e^+ + e^-$.

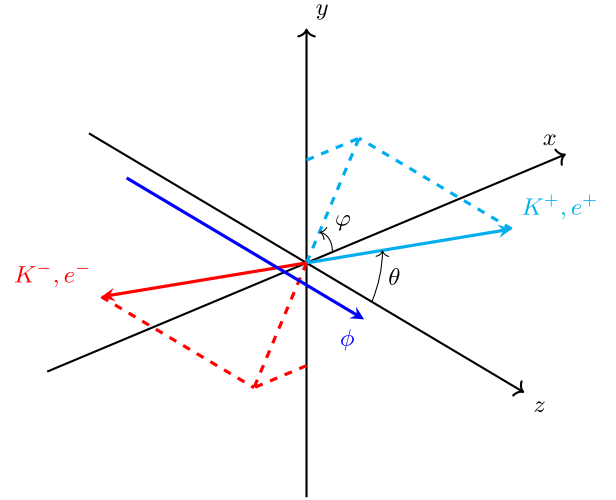

FIG. 2. Basic schematic of the $\phi$ meson decay in its rest frame. Cyan and red arrow each indicates daughter particle of positively charged and negatively charged. The blue arrow illustrates the flight direction of the $\phi$ meson in the Lab frame.

We assume that the initial vector meson state is a superposition of the polarization states $|\lambda\rangle$ ($\lambda = \pm 1$: transverse polarization, $\lambda = 0$: longitudinal polarization) with respective amplitudes $a_\lambda$. A general vector meson state can then be expressed as

$$|V\rangle = \sum_{\lambda=\pm 1, 0} a_\lambda |\lambda\rangle. \quad (2)$$

More specific definition of the four-polarization vector for each $\lambda$ is given in the Appendix A. Making use of the coefficients $a_\lambda$, we can define the spin density matrix $\rho_{\lambda\lambda'}$ without the bra-ket notation as

$$\rho_{\lambda\lambda'} = a_\lambda a_{\lambda'}^\star. \quad (3)$$

Note that the proper normalization of $|V\rangle$ ensures that the trace of the spin density matrix is normalized to unity; $\rho_{11} + \rho_{00} + \rho_{-1-1} = 1$. If the $\phi$ meson is transversely polarized, the z component of the $\phi$ meson spin will be $J_z = \pm 1$ and $\rho_{00} = 0$. On the other hand, if the $\phi$ meson is longitudinally polarized, $J_z = 0$ and $\rho_{00} = 1$. When the $\phi$ meson is unpolarized, diagonal entries of the density matrix are $\rho_{11} = \rho_{00} = \rho_{-1-1} = \frac{1}{3}$.

## A. $\phi \to K^+ + K^-$

The phenomenological interaction Lagrangian of the vector meson with two pseudoscalar mesons [17,19,20] is given as

$$\mathcal{L} = g_K \phi_\mu (K^+ \partial^\mu K^- - K^- \partial^\mu K^+), \quad (4)$$

where $\phi_\mu$ is the $\phi$ meson field and $K^+$, $K^-$ are charged kaon fields. For the evaluation of the phenomenological coupling between pseudoscalar mesons and the $\phi$ meson, denoted by $g_K$, we use the decay width of $\phi \to K^+ + K^-$ mentioned in the second line of Eq. (1) and the decay amplitude in Eq. (B4). Thus, we obtain $g_K = 4.476$ from the partial decay width of the $\phi$ meson. With the vector meson state assumed to be in the general configuration of Eq. (2), the angular distribution can be computed as [13]

$$\frac{1}{\Gamma}\frac{d\Gamma}{d\Omega} = \frac{3}{8\pi}(2\rho_{00}\cos^2\theta + (1-\rho_{00})\sin^2\theta$$
$$- 2\mathrm{Re}[\rho_{1-1}]\sin^2\theta\cos 2\varphi + 2\mathrm{Im}[\rho_{1-1}]\sin^2\theta\sin 2\varphi$$
$$- \sqrt{2}\mathrm{Re}[\rho_{10} - \rho_{-10}]\sin 2\theta \cos\varphi$$
$$+ \sqrt{2}\mathrm{Im}[\rho_{10} + \rho_{-10}]\sin 2\theta \sin\varphi), \quad (5)$$

where, $\theta$ and $\varphi$ are the polar and azimuthal angles of the outgoing $K^+$, and $|\mathbf{p}_1| = 127$ MeV the kaon momentum in the c.m. frame. $d\Gamma/d\Omega$ is obtained from Eq. (4). The details of this calculation are given in the Appendix B 2. Integrating $\frac{1}{\Gamma}\frac{d\Gamma}{d\Omega}$ over $\varphi$ from 0 to $2\pi$ yields the distributions $W(\theta)$ as

$$W(\theta) = \frac{3}{4}((1-\rho_{00}) + (3\rho_{00} - 1)\cos^2\theta). \quad (6)$$

In the approach followed in this paper, the coupling constants $g_K$ is determined by the strength of the kaonic decay with particles on their mass shells. Higher-order terms in the chiral expansion will lead to different vertices from that used in Eq. (4), with generally larger numbers of derivatives, which can change the analytic structure of the decay amplitude. However, whatever the form of the used vertex, the corresponding decay amplitude can always be cast in a form proportional to $F(m_\phi, m_K) \times \epsilon_\lambda \cdot (p_1 - p_2)$, because $p_2 = q - p_1$ and $\epsilon_\lambda \cdot q = 0$. The functional form of the coefficient $F(m_\phi, m_K)$ is determined by the details of the vertex, but will only depend on the $\phi$ and kaon masses, $m_\phi$ and $m_K$, because the incoming and outgoing particles are on the mass shell. As $F(m_\phi, m_K)$ is eventually determined from the decay width and $K^+ K^-$ branching ratio of the $\phi$, it can be absorbed into the definition of $g_K$. The angular distribution of the $K^+ K^-$ decay given in Eq. (5) therefore is not altered by higher-order chiral corrections.

Using a Lorentz boost, the angular dependence in the c.m. frame can be cast into the respective angular dependence in the Lab frame. If the initial vector meson momentum is higher than the momentum of the decaying particle in the c.m. frame (e.g., $q > 127$ MeV), the polar decay angle $\theta$ will have a maximum in the Lab frame. Furthermore, as a direct consequence, one $\theta$ value in the Lab frame will correspond to two polar-angle values in the c.m. frame. The decay amplitude is however uniquely determined if plotted as a function of the Lab-frame kaon momentum. More details regarding the relationship between the decay angles and momentum of the daughter particle in the Lab frame and the $\phi$ meson rest frame are worked out in the Appendix C.

Whatever the initial configuration may be $\rho_{\lambda\lambda'}$, it is generally not possible to remove all angular dependence of $\frac{1}{\Gamma}\frac{d\Gamma}{d\Omega}$ and make it completely isotropic. If, however, the azimuthal angle is not measured [or its average is taken as in Eq. (6)], all terms except the first two in Eq. (5) vanish, leading to the distribution $W(\theta)$, which can be independent of $\theta$. This happens if $\rho_{00} = 1/3$, in which case a vector meson is considered to be unpolarized. As can be recognized from Eq. (5), each polarization component has a different contribution at different angles. Specifically, the transverse components $(1 - \rho_{00}) = (\rho_{11} + \rho_{-1-1})$ contribute with a $\sin^2\theta$ term, while the longitudinal ($\rho_{00}$) with a $\cos^2\theta$ term on the right-hand side of Eq. (5). We can hence analyze how each component contributes and then look at specific angles to extract the relevant polarization component.

Figure 3 displays the decay rate of $\phi$ meson in the c.m. frame as a function of $\cos\theta$ for kaonic and electronic decays, respectively, in Figs. 3(a) and 3(b). Figures 4(a) and 4(b) show the decay rate of $\phi$ meson with given $\gamma\beta$ in the Lab frame, respectively, in terms of the $K^+$ and $e^+$ momenta in the Lab frame.

As can be seen in Fig. 3(a), the transverse component vanishes at $\theta = 0$ or $\pi$. This can readily be understood from the phenomenological Lagrangian of Eq. (4), which yields a hadronic decay amplitude proportional to the scalar product between the $K^+ K^-$ relative momentum and the

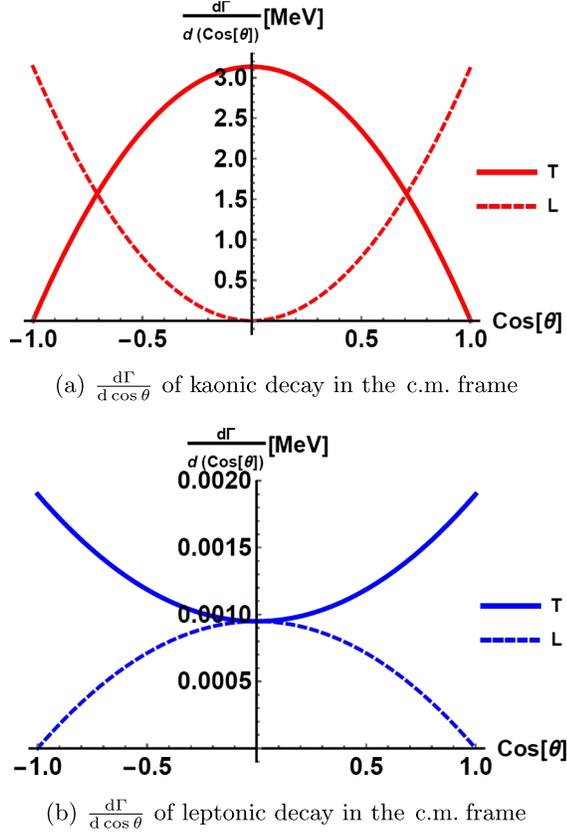

(a) $\frac{d\Gamma}{d\cos\theta}$ of kaonic decay in the c.m. frame

(b) $\frac{d\Gamma}{d\cos\theta}$ of leptonic decay in the c.m. frame

FIG. 3. Angular distribution of decay rate of (a) $\phi \to K^+ + K^-$ and (b) $\phi \to e^+ + e^-$ in the c.m. frame for each polarization. T stands for transverse polarization and L stands for longitudinal polarization.

polarization vector of the $\phi$ meson, $(\vec{p}_1 - \vec{p}_2) \cdot \vec{\epsilon}$. The details using polarization tensor of transverse and longitudinal modes satisfying the identity $P_{\mu\nu} = P_{\mu\nu}^T + P_{\mu\nu}^L = -g_{\mu\nu} + \frac{q_\mu q_\nu}{m_\phi^2}$ where $q_\mu$ is a four momentum of the $\phi$ meson, are discussed in the Appendix B 1. As a result, in the $\phi$ meson rest frame, the z component of the relative momentum contributes to the longitudinal amplitude, while the transverse momentum contributes to the transverse amplitude. Another, perhaps more intuitive, way of understanding the angular dependencies of the two polarization modes is to interpret the $K^+K^-$ decay amplitude as the orbital-angular momentum wave functions, which are proportional to the spherical harmonics, and behave as $Y_1^{\pm 1}(\theta, \varphi) \sim \sin\theta e^{\pm i\varphi}$ and $Y_1^0(\theta, \varphi) \sim \cos\theta$ for the transverse and longitudinal modes, respectively.

### B. $\phi \to e^+ + e^-$

The decay of the $\phi$ meson into dileptons can be obtained with the following interaction Lagrangians [16,17],

$$\mathcal{L}_{\phi\gamma} = -\frac{e}{2g_J} F^{\mu\nu} \phi_{\mu\nu}, \quad \mathcal{L}_{\gamma e^- e^+} = -e\bar{\psi}\gamma^\mu A_\mu \psi, \quad (7)$$

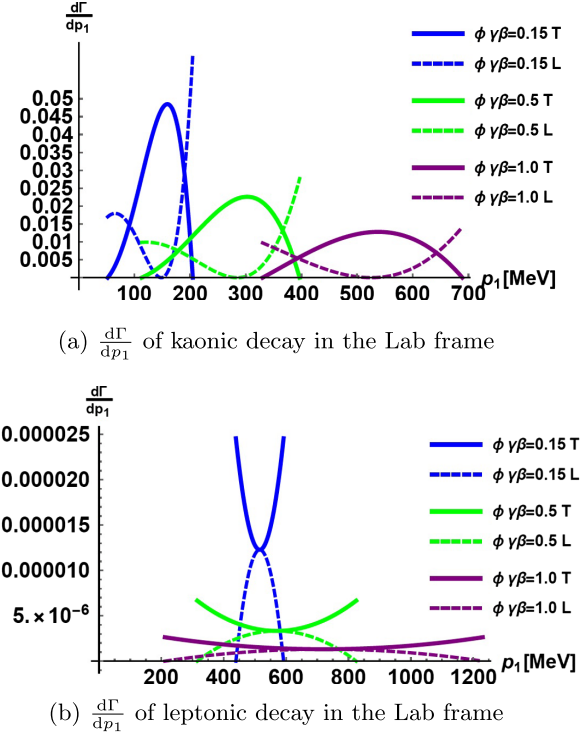

(a) $\frac{d\Gamma}{dp_1}$ of kaonic decay in the Lab frame

(b) $\frac{d\Gamma}{dp_1}$ of leptonic decay in the Lab frame

FIG. 4. Decay rate in the Lab frame for given $\gamma\beta$ of the $\phi$ meson as a function of momentum ($p_1$) of (a) $K^+$ and (b) $e^+$ measured from the Lab frame.

with, $F^{\mu\nu} = \partial^\mu A^\nu - \partial^\nu A^\mu$, $\phi_{\mu\nu} = \partial_\mu \phi_\nu - \partial_\nu \phi_\mu$, where $A^\mu$ is the photon field. $\mathcal{L}_{\phi\gamma}$ is the lowest-derivative term that is also gauge invariant. $g_J$ is the coupling constant between the $\phi$ meson and the photon and $\alpha = e^2/4\pi$ is the fine structure constant. The magnitude of the positron momentum in the c.m. frame is $|\boldsymbol{p}_1| = 509.73$ MeV. From the decay width of $\phi \to e^+ + e^-$, given in the third line of Eq. (1), we obtain $g_J = 13.4$.

Using Eq. (7), the angular distribution of the leptonic decay can be computed as follows:

$$\frac{1}{\Gamma}\frac{d\Gamma}{d\Omega} = \frac{3}{16\pi}(2(1-\rho_{00})\cos^2\theta + (1+\rho_{00})\sin^2\theta$$
$$+ 2\text{Re}[\rho_{1-1}]\sin^2\theta\cos 2\varphi - 2\text{Im}[\rho_{1-1}]\sin^2\theta\sin 2\varphi$$
$$+ \sqrt{2}\text{Re}[\rho_{10} - \rho_{-10}]\sin 2\theta\cos\varphi$$
$$- \sqrt{2}\text{Im}[\rho_{10} + \rho_{-10}]\sin 2\theta\sin\varphi), \quad (8)$$

where the same conventions as in the previous subsection were used. More details about the calculation of $d\Gamma/d\Omega$ from Eq. (7) are given in the Appendix B 3. As before, we can obtain $W(\theta)$ by integrating over $\varphi$, leading to

$$W(\theta) = \frac{3}{8}((1+\rho_{00}) + (1-3\rho_{00})\cos^2\theta). \quad (9)$$

The transverse and longitudinal amplitudes for the leptonic decay display a different angular dependence to that of the $K^+K^-$ decay. Specifically, the longitudinal amplitude vanishes in the forward ($\theta = 0$) or backward ($\theta = \pi$) direction, as shown in Fig. 3(b). This can be understood from the way the angular momentum of the $\phi$ meson is passed over to the dilepton spins. The $\phi$ decays into dileptons via a coupling to a photon, as expressed in Eq. (7), and the electromagnetic photon-dilepton coupling allows only dilepton pairs with opposite helicity, which should be conserved for (approximately) massless dileptons. Therefore, for forward or backward decays, the dilepton spins must be either pointing in the forward or backward direction, and hence do not couple to the initial longitudinal $\phi$ meson mode. Detailed calculations showing how this works when making use of explicit electron and positron spinors are given in the Appendix B 4.

One can also understand the difference between the kaonic and dilepton decays from the difference of the tensor structures appearing in Eqs. (4) and (7). For the electronic decay, the relative strength between the different tensor structures in the lower expression of Eq. (B2), which is controlled by the photon-electron coupling and determines the angular dependence in the decay, is uniquely determined by requiring it to vanish after contracting by both $q^\mu$ and $k^\mu = (p_1^\mu - p_2^\mu)$. The first constraint is due to the Ward identity (and is hence exact), while the second is true only to leading order in $\alpha$ because at higher order a $\sigma_{\mu\nu}/2m$-type correction appears in the electron-photon vertex, which is related to the anomalous magnetic moment of the electron. As this is only a higher-order QED effect, we can safely neglect it here. For the hadronic vertex $F_{\mu\nu}\phi^{\mu\nu}$ in Eq. (7), there will again be higher-order terms with more (covariant) derivatives. Following the same argument as for the kaonic decay above, these will only affect the overall constant of the decay amplitude, can thus be absorbed into the definition of $g_J$ and will not affect the angular distribution of the $e^+e^-$ decay.

### C. Unified treatment using the Wigner D-matrix

In the above subsections, we have calculated the general angular distribution by using the respective interaction Lagrangian. Let us briefly remark here that, stemming from the fact that the $\phi$ meson carries angular momentum 1, one can reproduce the results of both decay channels using the Wigner D-matrix for spin-1 states, for which the convention identical to that of Ref. [21] was adopted. Specifically, we need the matrix

$$\mathcal{D}^1_{mm'}(\varphi,\theta,-\varphi)$$
$$= e^{-im\varphi}d^1_{mm'}(\theta)e^{im'\varphi}$$
$$= \begin{pmatrix} \frac{1+\cos\theta}{2} & -\frac{1}{\sqrt{2}}\sin\theta e^{-i\varphi} & \frac{1-\cos\theta}{2}e^{-2i\varphi} \\ \frac{1}{\sqrt{2}}\sin\theta e^{i\varphi} & \cos\theta & -\frac{1}{\sqrt{2}}\sin\theta e^{-i\varphi} \\ \frac{1-\cos\theta}{2}e^{2i\varphi} & \frac{1}{\sqrt{2}}\sin\theta e^{i\varphi} & \frac{1+\cos\theta}{2} \end{pmatrix}, \quad (10)$$

to rotate $\rho_{mm'}$, for which the z axis is the quantization axis. The above rotation brings the quantization axis to align with the outgoing $e^+$ or $K^+$. For kaon decay, only a longitudinally-polarized $\phi$ meson can decay into $K^+$ and $K^-$ in the forward or backward direction as mentioned in Sec. II A,

$$\left(\frac{1}{\Gamma}\frac{d\Gamma}{d\Omega}\right)_{K^+K^-} \propto \sum_{m,m'=\pm 1,0} D^{1\dagger}_{0m}\rho_{mm'}D^1_{m'0}, \quad (11)$$

where only the 00 matrix element of the rotated spin density matrix contributes.

For the dilepton decay, the $\phi$ meson in contrast couples to a forwardly decaying dilepton only if it is transversely polarized. Therefore,

$$\left(\frac{1}{\Gamma}\frac{d\Gamma}{d\Omega}\right)_{e^+e^-} \propto \sum_{\lambda=\pm 1}\sum_{m,m'=\pm 1,0} D^{1\dagger}_{\lambda m}\rho_{mm'}D^1_{m'\lambda}, \quad (12)$$

where $\lambda = 1$ and $\lambda = -1$ contribute with equal weight due to parity conservation.

### III. DISCUSSIONS

As we have seen in the above section, $K^+K^-$ decay and $e^+e^-$ decay channels show a different behavior in terms of angular dependence of their decay amplitudes. This fact can help to discern transverse and longitudinal polarization from an unpolarized $\phi$ meson.

The lepton channel is at first sight advantageous because of its small final state interaction. However, the J-PARC E16 experiment has limited acceptance for $\phi \to e^+e^-$ at $\cos(\theta) = \pm 1$, where 100% transverse polarization is expected. In case of $\cos(\theta) = \pm 1$, either electron or positron is emitted opposite to the $\phi$ meson in its rest frame. The boost to the lab frame makes the backward lepton momentum very low. E16 has small acceptance for lepton with a momentum of $< 0.4$ GeV/c because of a minimum energy requirement that are needed online to suppress the background, and a large bending angle in the magnetic field of E16 spectrometer.

Since J-PARC E88 will have almost uniform acceptance for $K^\pm$ angles due to their small opening angles and the Lorentz boost in the laboratory frame, we can extract purely longitudinally-polarized $\phi$ at $\theta = 0°$ and $180°$ and purely transversely-polarized $\phi$ at $\theta = 90°$. However, we remark that the $K^\pm$ decay angle is distorted due to final state interactions between $K^\pm$ and nucleons, due to the strong interaction.

### IV. SUMMARY AND CONCLUSIONS

We have in this paper discussed how the $\phi$ meson in a general polarization state decays into $e^+e^-$ and $K^+K^-$ decay channels. We showed that the angular dependencies of the respective decay amplitudes can be derived both from the explicit Lagrangians of Eqs. (4) and (7), but also from simple considerations involving angular-momentum conservation and properties of the spin density matrix under rotations.

Furthermore, we examined the feasibility of experimentally disentangling the longitudinal and transverse polarization modes of the $\phi$ meson and of measuring their individual in-medium properties at the J-PARC E16 and E88 experiments. We find that both $e^+e^-$ and $K^+K^-$ channels have their advantages and disadvantages. While $e^+e^-$ would be preferred because it does not suffer from strong final-state interactions, it will be difficult to measure forward (or backward) decaying electrons which would be the most useful configuration to measure the two polarization modes separately. $K^+K^-$, on the other hand, will suffer from distortions due to strong $KN$ and $\bar{K}N$ interactions in the final state, but a full and approximately uniform coverage of all decay angles will be possible. All this suggests that a measurement of both $e^+e^-$ and $K^+K^-$ decays may be useful as the two decays can provide complementary information.

### ACKNOWLEDGMENTS

This work was supported by Samsung Science and Technology Foundation under Project No. SSTF-BA1901-04, JSPS KAKENHI Grants No. JP19KK0077, JP20K03940, JP21H00128, and JP21H01102, and the Leading Initiative for Excellent Young Researchers (LEADER) of the Japan Society for the Promotion of Science (JSPS).

### APPENDIX A: SPIN, POLARIZATION AND MOMENTUM OF A VECTOR MESON

We perform our calculation of the angular distribution of the $\phi$ meson decay amplitude using circularly polarized bases consisting of two transverse and one longitudinal polarization vector. Note that the discussion here can be applied to any vector meson with a nonzero mass. Any vector meson state with finite mass can be defined as superposition of three different polarization states.

A spin-1 particle can be described by the vector representation of the SO(3) group. Generators and commutation relations of a rotation are given as $(S_i)_{jk} = -i\epsilon_{ijk}$, $[S_i, S_j] = i\epsilon_{ijk} S_k$ $(i,j = 1, 2, 3)$, with

$$S_1 = \begin{pmatrix} 0 & 0 & 0 \\ 0 & 0 & -i \\ 0 & i & 0 \end{pmatrix}, \quad S_2 = \begin{pmatrix} 0 & 0 & i \\ 0 & 0 & 0 \\ -i & 0 & 0 \end{pmatrix},$$

$$S_3 = \begin{pmatrix} 0 & -i & 0 \\ i & 0 & 0 \\ 0 & 0 & 0 \end{pmatrix}. \quad (A1)$$

The three eigenvectors of $S_3$ constitute spatial part of circularly polarized bases of $S_3$ in the $\phi$ meson rest frame. $\vec{\varepsilon}_{\pm 1}$ are transversely polarized eigenstates corresponding to $S_3 = \pm 1$ and $\vec{\varepsilon}_0$ is longitudinally polarized eigenstate of $S_3 = 0$. Vector-meson state and four-momentum are also given below.

$$\vec{\varepsilon}_{\pm 1} = \left(\mp \frac{1}{\sqrt{2}}, -\frac{i}{\sqrt{2}}, 0\right), \qquad \vec{\varepsilon}_0 = (0, 0, 1),$$

$$|\pm 1\rangle = (0, \vec{\varepsilon}_{\pm 1}), \qquad |0\rangle = (0, 0, 0, 1),$$

$$|V\rangle = \sum_{\lambda=\pm 1,0} a_\lambda |\lambda\rangle, \qquad q^\mu = (m_\phi, 0, 0, 0).$$

### APPENDIX B: COMPUTATION OF THE DECAY AMPLITUDE ANGULAR DEPENDENCE

In this appendix, it will be demonstrated in some detail how to obtain the general angular of the $e^+e^-$ and $K^+K^-$ decay amplitudes.

#### 1. Separating polarization modes of a spin-1 particle

Let us start with the simple case for the decay of a purely longitudinally or transversely-polarized state. To separate the transverse and longitudinal part in components of the decay amplitude, it is convenient to apply the projection operators constructed with the help of the polarization vectors. Using the four-momentum $q^\mu$ and polarization vectors $\varepsilon^\mu$ given in the Appendix A, we obtain the transverse and longitudinal projection operators as

$$P^T_{\mu\nu} = \varepsilon^T_\mu \varepsilon^{T\star}_\nu = \begin{pmatrix} 0 & 0 \\ 0 & \delta_{ij} - \frac{q_i q_j}{\vec{q}^2} \end{pmatrix},$$

$$P^L_{\mu\nu} = \varepsilon^L_\mu \varepsilon^{L\star}_\nu = \begin{pmatrix} \frac{\vec{q}^2}{m_\phi^2} & -\frac{q_0 q_i}{m_\phi^2} \\ -\frac{q_0 q_i}{m_\phi^2} & \frac{q_0^2 q_i q_j}{m_\phi^2 \vec{q}^2} \end{pmatrix}. \quad (B1)$$

Here, superscripts $T$ and $L$ stand for transverse and longitudinal parts, respectively. $q_0$ and $q_i$ are energy and momentum of the vector (here, $\phi$) meson, $i$ and $j$ run from 1 to 3 and $q_i = 0$ in the vector-meson rest frame. To calculate the decay amplitude for each polarization mode, we need the respective $\phi \to K^+ + K^-$ and $\phi \to e^+ + e^-$ decay amplitudes (with unspecified polarization of the initial $\phi$ meson). They can be derived from Eqs. (4) and (7) as

$$\mathcal{M}^{\mu\nu}_{\text{Hadronic}} = g_K^2 (p_1 - p_2)^\mu (p_1 - p_2)^\nu,$$

$$\mathcal{M}^{\mu\nu}_{\text{Leptonic}} = \frac{64\pi^2 \alpha^2}{g_J^2} \left( p_1^\mu p_2^\nu + p_2^\mu p_1^\nu - \frac{1}{2} m_\phi^2 g^{\mu\nu} \right). \quad (B2)$$

Multiplying the projection operators of Eq. (B1) with the above decay amplitudes and tracing the Lorentz indices yields the decay amplitude for each polarization. Specifically, we obtain the results as compiled below.

$$\phi \to K^+ + K^- \begin{cases} |\mathcal{M}|^2_T = 2g_K^2 |\boldsymbol{p_1}|^2 \sin^2\theta, \\ |\mathcal{M}|^2_L = 4g_K^2 |\boldsymbol{p_1}|^2 \cos^2\theta, \end{cases}$$

$$\phi \to e^+ + e^- \begin{cases} |\mathcal{M}|^2_T = \frac{64\pi^2\alpha^2}{g_J^2} (\frac{1}{2}m_\phi^2 - |\boldsymbol{p_1}|^2 \sin^2\theta), \\ |\mathcal{M}|^2_L = \frac{64\pi^2\alpha^2}{g_J^2} (\frac{1}{2}m_\phi^2 - 2|\boldsymbol{p_1}|^2 \cos^2\theta). \end{cases} \quad (B3)$$

### 2. General angular distribution of the $\phi \to K^+ + K^-$ decay

Next, we consider the more general case, in which the initial vector meson is in a linear superposition of all possible spin configurations, as expressed in Eq. (2). As before, the $\phi \to K^+ + K^-$ decay amplitude follows from the interaction Lagrangian given in Eq. (4). We have

$$\mathcal{M} = g_K (p_1 - p_2)^\mu \sum_{\lambda=\pm 1,0} a_\lambda \varepsilon_{\lambda\mu}, \quad (B4)$$

where $p_1 = (E_1, \boldsymbol{p_1})$ and $p_2 = (E_2, \boldsymbol{p_2})$ are the four-vectors of the emitted kaons, with $|\boldsymbol{p_1}| = |\boldsymbol{p_2}| = \frac{1}{2}\sqrt{m_\phi^2 - 4m_K^2}$, where $m_K$ is the kaon mass. Taking the absolute square and summing over the initial $\phi$ meson polarizations, yields the general angular distribution of this decay process as

$$|\mathcal{M}|^2 = 2g_K^2 |\boldsymbol{p_1}|^2 (2\rho_{00}^2 \cos^2\theta + (1 - \rho_{00})\sin^2\theta \\ - 2\text{Re}[\rho_{1-1}]\sin^2\theta \cos 2\varphi + 2\text{Im}[\rho_{1-1}]\sin^2\theta \sin 2\varphi \\ - \sqrt{2}\text{Re}[\rho_{10} - \rho_{-10}] \sin 2\theta \cos\varphi \\ + \sqrt{2}\text{Im}[\rho_{10} + \rho_{-10}] \sin 2\theta \sin\varphi). \quad (B5)$$

### 3. General angular distribution of the $\phi \to e^+ + e^-$ decay

The $\phi \to \gamma^* \to e^+ + e^-$ decay amplitude can be calculated using the interaction Lagrangian in Eq. (7). We obtain

$$\mathcal{M} = \frac{4\pi\alpha}{g_J} \bar{u}(p_2) \gamma^\mu \sum_{\lambda=\pm 1,0} a_\lambda \varepsilon_{\lambda\mu} v(p_1), \quad (B6)$$

where, as before, $p_1$ and $p_2$ are the four-vectors of the decaying dileptons, while $v(p_1)$ and $u(p_2)$ are the Dirac spinors of the positron and electron, respectively. The magnitude of the spatial components of $p_1$ and $p_2$, can be given as $|\boldsymbol{p_1}| = |\boldsymbol{p_2}| = \frac{1}{2}\sqrt{m_\phi^2 - 4m_l^2} \approx \frac{1}{2}m_\phi$, where $m_l$ is the electron mass, that can be safely ignored here. Repeating the procedure of the previous subsection, we can obtain the general angular distribution as

$$|\mathcal{M}|^2 = \frac{16\pi^2\alpha^2 m_\phi^2}{g_J^2} (2(1 - \rho_{00})\cos^2\theta + (1 + \rho_{00})\sin^2\theta \\ + 2\text{Re}[\rho_{1-1}]\sin^2\theta \cos 2\varphi - 2\text{Im}[\rho_{1-1}]\sin^2\theta \sin 2\varphi \\ + \sqrt{2}\text{Re}[\rho_{10} - \rho_{-10}] \sin 2\theta \cos\varphi \\ - \sqrt{2}\text{Im}[\rho_{10} + \rho_{-10}] \sin 2\theta \sin\varphi). \quad (B7)$$

### 4. Helicity of $e^+e^-$ states generated by the vector current $\bar{u}(p_2)\gamma^\mu v(p_1)$

For an intuitive understanding of the $e^+e^-$ decay amplitude, it is instructive to study the helicity structure of the outgoing electron and positron generated by the vector current $\bar{u}(p_2)\gamma^\mu v(p_1)$. For this purpose, we use explicit Dirac spinors in the $\phi$-meson rest frame. In analogy to the preceding subsections, $p_1$ and $p_2$ are the positron and electron momentum where $|\boldsymbol{p_1}| = |\boldsymbol{p_2}|$ and $E_1$ and $E_2$ are the positron and electron energies, respectively. In the $\phi$ meson rest frame, they are equal and will here hence be denoted as $E_1 = E_2 = E$. We use the chiral representation for the positron and electron spinors, and thus have $\gamma^\mu = (\gamma_0, \gamma_i)$ with

$$\gamma_0 = \begin{pmatrix} 0 & 1 \\ 1 & 0 \end{pmatrix}, \quad \gamma_i = \begin{pmatrix} 0 & \sigma_i \\ -\sigma_i & 0 \end{pmatrix}, \quad \gamma_5 = \begin{pmatrix} -1 & 0 \\ 0 & 1 \end{pmatrix}. \quad (B8)$$

Let us consider here the case, in which the positron (electron) is emitted into the forward (backward) direction. This means that their spinors become eigenstates of the spin operator with respect to the $z$-axis, which in the limit of vanishing electron mass can be explicitly given as

$$v^1(p_1) = \sqrt{2E_1}\begin{pmatrix} 0 \\ 1 \\ 0 \\ 0 \end{pmatrix}, \quad v^2(p_1) = \sqrt{2E_1}\begin{pmatrix} 0 \\ 0 \\ -1 \\ 0 \end{pmatrix},$$

$$u^1(p_2) = \sqrt{2E_2}\begin{pmatrix} 1 \\ 0 \\ 0 \\ 0 \end{pmatrix}, \quad u^2(p_2) = \sqrt{2E_2}\begin{pmatrix} 0 \\ 0 \\ 0 \\ 1 \end{pmatrix}. \quad (B9)$$

Using the above spinors, we can compute the vector current for each combination of helicity as

$$\bar{u}^1 \gamma^\mu v^1 = (0, -2E, 2iE, 0), \quad \bar{u}^1 \gamma^\mu v^2 = (0, 0, 0, 0),$$
$$\bar{u}^2 \gamma^\mu v^1 = (0, 0, 0, 0), \quad \bar{u}^2 \gamma^\mu v^2 = (0, -2E, -2iE, 0). \quad (B10)$$

To go further and compute the vector current for a general emission angle, we simply have to rotate the above result. Specifically, we should apply subsequent rotations around the $y$-axis by $\theta$ and thereafter around the $z$-axis by $\varphi$,

$$\begin{pmatrix} 1 & 0 & 0 & 0 \\ 0 & \cos\varphi & -\sin\varphi & 0 \\ 0 & \sin\varphi & \cos\varphi & 0 \\ 0 & 0 & 0 & 1 \end{pmatrix} \cdot \begin{pmatrix} 1 & 0 & 0 & 0 \\ 0 & \cos\theta & 0 & \sin\theta \\ 0 & 0 & 1 & 0 \\ 0 & -\sin\theta & 0 & \cos\theta \end{pmatrix}, \quad (B11)$$

which leads to the following nonvanishing vector currents without cross terms with opposite helicity:

$$\bar{u}^1(p_2)\gamma^\mu v^1(p_1) = 2E(0, -\cos\theta\cos\varphi - i\sin\varphi, -\cos\theta\sin\varphi + i\cos\varphi, \sin\theta),$$
$$u^2(p_2)\gamma^\mu v^2(p_1) = 2E(0, -\cos\theta\cos\varphi + i\sin\varphi, -\cos\theta\sin\varphi - i\cos\varphi, \sin\theta). \quad (B12)$$

## APPENDIX C: RELATIONSHIP BETWEEN DECAY ANGLE MEASURED IN THE $\phi$-MESON REST FRAME AND THE LAB FRAME

The kinematics of the decays of interest here in the Lab and $\phi$-meson rest frame are related via a Lorentz transformation. Its transverse and longitudinal components can be written down as

$$p_1 \sin\xi = p_1' \sin\theta,$$
$$p_1 \cos\xi = \gamma(p_1' \cos\theta + \beta E_1'),$$
$$p_1' \cos\theta = \gamma(p_1 \cos\xi - \beta E_1), \quad (C1)$$

where $\xi$ and $\theta$ stand for the polar angles of the positively charged particle with respect to the $\phi$-meson momentum in the Lab and rest frame. Likewise, $p_1$, $E_1$ and $p_1'$, $E_1'$ are the momentum and energy in the Lab frame and the $\phi$ meson rest frame, respectively. $\beta$ and $\gamma$ are the velocity of the $\phi$ meson in the Lab frame and the corresponding Lorentz factor.

Using Eq. (C1), one can straightforwardly relate quantities in the Lab frame to those in the rest frame. As an example, the Lab-frame polar angle $\xi$ is depicted in Fig. 5 as a function of $\theta$ for three representative $\beta\gamma$ values. It is seen there that the range of $\xi$ depends on the velocity of the $\phi$ meson. If $\beta < \frac{p_1'}{E_1'}$ ($\frac{p_1'}{E_1'} = 0.249$ for $\phi \to K^+K^-$), $\xi$ is a monotonic increasing function of $\theta$. On the other hand, if $\beta > \frac{p_1'}{E_1'}$, the mapping from $\theta$ to $\xi$ becomes two to one,

$$\beta < \frac{p_1'}{E_1'}, \quad 0 \leq \xi \leq \pi,$$
$$\beta > \frac{p_1'}{E_1'}, \quad 0 \leq \xi \leq \xi_{\max}. \quad (C2)$$

We can furthermore determine the Lab frame momentum of $K^+/e^+$ as a function of the Lab frame angle $\xi$ as

$$p_{1\pm} = \frac{\sqrt{\gamma^2 E_1'^2 - m_1^2\cos^2\xi + (\gamma^2-1)p_1'^2\cos^2\xi - \gamma^4 m_1^2\sin^2\xi} \pm 2\gamma^2\beta E_1'\sqrt{\cos^2\xi(p_1'^2 - \gamma^2\beta^2 m_1^2\sin^2\xi)}}{\gamma^2(1-\beta^2\cos^2\xi)}, \quad (C3)$$

where $m_1$ is the mass of the decay particles.

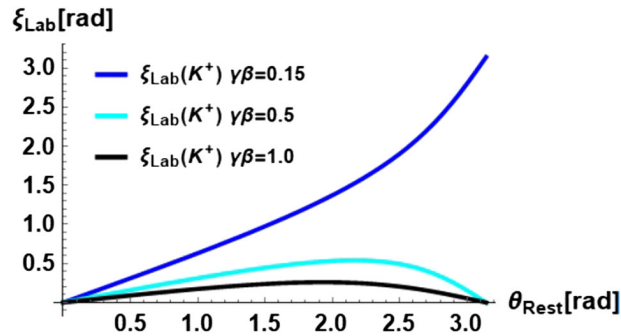

FIG. 5. Relationship between decay angle measured in the Lab frame ($\xi$) and the $\phi$-meson rest frame ($\theta$) in kaonic decay.